\RequirePackage{fix-cm}
\documentclass[smallextended]{svjour3}       
\smartqed  
\usepackage{graphicx}
\journalname{Eur. Phys. J. Plus}
\usepackage{epsfig}
\usepackage{amsfonts,amssymb}
\usepackage{color}

  \usepackage{slashed}
  \usepackage{url}
\usepackage{bm}
\usepackage{verbatim}
\usepackage{float}

\begin{document}

\title{Quantum Yang-Mills field theory
}


\author{Marco Frasca         
}


\institute{Marco Frasca \at
              Via Erasmo Gattamelata, 3 \\
							00176 Rome (Italy)
              \email{marcofrasca@mclink.it}           
}

\date{Received: date / Accepted: date}

\maketitle

\begin{abstract}
We show that the Dyson-Schwinger set of equations for the Yang-Mills theory can be exactly solved till the two-point function.
This is obtained given a set of nonlinear waves solving the classical equations of motion.
Translation invariance is maintained by the proper choice of the solution of the equation for the two-point function as devised by Coleman. The computation of the Dyson-Schwinger equations is performed in the same way as devised by Bender, Milton and Savage providing a set of partial differential equations whose proof of existence of the solutions is standard. So, the correlation functions of the theory could be proved to exist and the two-point function manifests a mass gap.
\keywords{Yang-Mills theory \and Scalar field theory \and Dyson-Schwinger equations \and Mass spectrum}
\end{abstract}

\section{Introduction}

Classical solutions of nonlinear field theories has been known for long. Typically one gets solitons or wave-like solutions. The question on how to build a quantum field theory starting from these solutions was assessed on '70s \cite{Cornwall:1974vz,Vasilev:1974ze,Goldstone:1974gf,Tomboulis:1975gf,Klein:1975yh,Jacobs:1975zh,Banks:1975zw}. A recent review has been given in \cite{Weinberg:2012pjx}. It was pointed out that the chosen classical solutions should be stable with respect to the classical energy functional. This boils down to solve an eigenvalue problem.

Dyson-Schwinger equations \cite{Dyson:1949ha,Schwinger:1951ex,Schwinger:1951hq} represent a powerful approach to treat a quantum field theory in a non-perturbative way. Difficulties arise due to the fact that the lower order equations depends on higher order correlation functions that are solutions to higher order equations. One way out to this dilemma is to truncate the hierarchy and neglect correlation functions beyond a given order. The main problem with such an approach is that one does not know if the truncation choice is a correct one and the chosen approximation is generally out of control, differently from perturbation techniques.

Dyson-Schwinger equation for Yang-Mills theory are generally treated as coupled integral equations in momenta space \cite{Mandelstam:1979xd,Atkinson:1981er,Atkinson:1981ah,Brown:1988bn,vonSmekal:1997vx}. The preferred formulation is in the Landau gauge that makes computations easier. We will follow the same line. The main difference in our paper is the choice of the way to derive the Dyson-Schwinger equations. We will follow the approach devised in Ref.\cite{Bender:1999ek} by Bender, Milton and Savage. The reason to do this relies on the fact that working with partial differential equations is simpler in this case as we know their exact solutions \cite{Frasca:2009bc,Frasca:2014xaa}.

The aim of this paper is to show that, given a set of classical solutions, for this particular set all the hierarchy of Dyson-Schwinger equations could be, in principle, exactly solved. This is seen extensively for a scalar field theory and for the Yang-Mills theory just till the two-point function. The set of solutions is proven to be stable in agreement with the studies in the '70s and the way how renormalization comes in is explicitly shown for the scalar field theory. Both theories display a mass gap. The relevance of the chosen set of classical solutions to start with relies on what is effectively observed experimentally. In any case, it is interesting to see that such a possibility exists for the difficult case of Dyson-Schwinger equations.

An interesting point arises due to the nonlinear nature of the differential equations we obtain. In this case it appears as the invariance by translation cannot hold and all the procedure is doomed. The way out to this problem has been devised long ago by Sidney Coleman \cite{Coleman:1985}. The idea is to take the derivative of the one-point function to solve for the two-point function. Then, this solution is invariant by construction.
The fact that the translation invariance is preserved is also seen by the existence of a zero mode in the eigenvalue problem.

In this paper we are able to show that, in this way, the Dyson-Schwinger hierarchy for the Yang-Mills quantum field theory can be exactly solved at least till the two-point function in the Landau gauge. The proof of existence for correlation functions are easier to obtain working with partial differential equations. Besides, the two-point function displays a mass gap.


It should be pointed out that, in our case, mass generation is granted by a proper choice of the boundary conditions that are normally extended to an infinite volume. Otherwise, the mechanism cannot be granted to generate mass as pointed out both for the case of the scalar field \cite{Nogueira:2001gg,Sissakian:2004aj} and Coleman-Weinberg mechanism \cite{Fagundes:2012xu,Fagundes:2010zzb}. About confinement, it should be emphasized that it is the combination of Yukawa-like propagators, showing mass gap, and a running coupling to possibly yield a confining potential \cite{Gonzalez:2011zc,Vento:2012wp,Deur:2016bwq}. We assume this as a working mechanism without further proof in this article.

The paper is so structure, in Sec.~\ref{sec2} we discuss the case of the scalar field theory. This will display the way the technique works and the solutions we obtain will be fundamental to understand the case of the Yang-Mills theory. In Sec.~\ref{sec2a} we analyze the renormalization conditions for the solution. In Sec.~\ref{sec2b}we prove that the classical solutions we choose to start with are indeed stable. In Sec.~\ref{sec3} we yield the exact solutions of the Yang-Mills classical field equations. Finally, in Sec.~\ref{sec4} we solve the Dyson-Schwinger hierarchy till the two-point function for the Yang-Mills theory. In Sec.~\ref{sec5} conclusions are given.

\section{Scalar field}
\label{sec2}

We want to solve the quantum equation of motion
\begin{equation}
    \partial^2\phi+\lambda\phi^3=j
\end{equation}
given the generating functional
\begin{equation}
    Z[j]=\int[d\phi]e^{i\int d^4x\left[\frac{1}{2}(\partial\phi)^2-\frac{\lambda}{4}\phi^4+j\phi\right]}.
\end{equation}
We average on the vacuum state $|0\rangle$ and divide by $Z[j]$ yielding
\begin{equation}
\label{eq:sf1}
    \partial^2 G_1^{(j)}(x)+\lambda\frac{\langle 0|\phi^3|0\rangle}{Z[j]}=j
\end{equation}
where we have defined $G_1^{(j)}(x)=\langle 0|\phi|0\rangle/Z[j]$, the one-point function. Then we write
\begin{equation}
    G_1^{(j)}(x)Z[j]=\langle 0|\phi|0\rangle
\end{equation}
and we take the functional derivative with respect to $j$ obtaining
\begin{equation}
    [G_1^{(j)}(x)]^2Z[j]+G_2^{(j)}(x,x)Z[j]=\langle 0|\phi^2|0\rangle
\end{equation}
and deriving again one has
\begin{equation}
    [G_1^{(j)}(x)]^3Z[j]+3G_2^{(j)}(x,x)G_1^{(j)}(x)Z[j]+G_3^{(j)}(x,x,x)Z[j]=\langle 0|\phi^3|0\rangle.
\end{equation}
Using eq.(\ref{eq:sf1}), this becomes
\begin{equation}
\label{eq:g1}
   \partial^2 G_1^{(j)}(x)+\lambda\left([G_1^{(j)}(x)]^3+3G_2^{(j)}(x,x)G_1^{(j)}(x)+G_3^{(j)}(x,x,x)\right)=j.
\end{equation}
Taking $j=0$, observing that the theory is invariant by translations, that is $G_2(x,y)=G_2(x-y)$, one has the first Dyson-Schwinger equation of the scalar theory
\begin{equation}
\label{eq:g10}
   \partial^2 G_1(x)+\lambda\left([G_1(x)]^3+3G_2(0)G_1(x)+G_3(0,0)\right)=0.
\end{equation}
This equation can be solved exactly as we show in a moment, respecting translation invariance. Before to see this, we derive the Dyson-Schwinger equation for the two-point function. We take the functional derivative of eq.(\ref{eq:g1}) to obtain
\begin{eqnarray}
   &&\partial^2G_2^{(j)}(x,y)+\lambda\left(3[G_1^{(j)}(x)]^2G_2^{(j)}(x,y)+3G_2^{(j)}(x,x)G_2^{(j)}(x,y)\right. \nonumber \\
	&&\left.+3G_3^{(j)}(x,x,y)G_1^{(j)}(x)+G_4^{(j)}(x,x,x,y)\right)=\delta^4(x-y).
\end{eqnarray}
Taking $j=0$ we finally obtain
\begin{eqnarray}
\label{eq:g20}
   &&\partial^2G_2(x-y)+\lambda\left(3[G_1(x)]^2G_2(x-y)+3G_2(0)G_2(x-y)\right. \nonumber \\
	&&\left.+3G_3(0,x-y)G_1(x)+G_4(0,0,x-y)\right)=\delta^4(x-y).
\end{eqnarray}
This equation, written in this way, seems to break translation invariance due to $G_1(x)$. This is so unless we choose properly the solution of the one-point function. Indeed, the solution to eq.(\ref{eq:g10}) can be written as
\begin{equation}
    G_1(x)=\sqrt{\frac{2\mu^4}{m^2+\sqrt{m^4+2\lambda\mu^4}}}{\rm sn}\left(p\cdot x+\chi,\frac{-m^2+\sqrt{m^4+2\lambda\mu^4}}{-m^2-\sqrt{m^4+2\lambda\mu^4}}\right)
\end{equation}
being $\mu$ and $\chi$ arbitrary integration constants and provided we put $m^2=3\lambda G_2(0)$, $G_3(0,0)=0$ and take the momenta $p$ so that
\begin{equation}
\label{eq:disp}
    p^2=m^2+\frac{\lambda\mu^4}{m^2+\sqrt{m^4+2\lambda\mu^4}}.
\end{equation}
It is important to notice that $G_2(0)$ gives a mass correction while the dispersion relation would be massive even if $G_2(0)=0$. We will evaluate this correction later. The main point is how to fix the arbitrary phase $\theta$. 
This phase, together with the parameter $\mu$, shows that the number of solutions for a Dyson-Schwinger set of equations is infinite and boundary condition should properly given to fix one \cite{Bender:1988bp,Guralnik:1996ri}. This will be seen below.
%
In any case, it
must be done consistently throughout all the tower of Dyson-Schwinger equations to preserve translation invariance. For eq.(\ref{eq:g20}) this is accomplished straightforwardly by taking $\chi=-p\cdot y+\chi'$. Then one has, making explicit the dependence on $y$ in $G_1$.
\begin{eqnarray}
\label{eq:g200}
   &&\partial^2G_2(x-y)+\lambda\left(3[G_1(x-y)]^2G_2(x-y)\right. \nonumber \\
	&&+3G_2(0)G_2(x-y)+3G_3(0,x-y)G_1(x-y) \nonumber \\
	&&\left.+G_4(0,0,x-y)\right)=\delta^4(x-y).
\end{eqnarray}
and we get a consistent Dyson-Schwinger equation with respect to the symmetries of the theory. 
In order to get the two-point function we determine the solutions of the equation
\begin{equation}
\label{eq:y200}
   \partial^2w(x)+\lambda\left(3[G_1(x)]^2w(x)+3G_2(0)w(x)\right)=0.
\end{equation}
This can be rewritten as
\begin{equation}
\label{eq:y201}
   \partial^2w(x)+m^2w(x)+3\lambda[G_1(x)]^2w(x)=0.
\end{equation}
Our solution must preserve translation invariance and so \cite{Coleman:1985}
\begin{equation}
   w_1(\zeta)=\frac{dG_1(\zeta)}{d\zeta}
\end{equation}
having set $\zeta=p\cdot x+\chi$. The other independent solution has the form
\begin{equation}
   w_2(\zeta)=\frac{1}{4}\zeta\frac{dG_1(\zeta)}{d\zeta}+\frac{1}{2}G_1(\zeta).
\end{equation}
$w_1(\zeta)$ represents the zero mode of the theory, this is the eigenfunction with 0 eigenvalue. The partition function will become infinite and one has a symmetry breaking. We consider this solution to determine the two-point function as the other breaks translation invariance. We will have
\begin{equation}
   w_1(\zeta)=\frac{dG_1(\zeta)}{d\zeta}=\sqrt{\frac{2\mu^4}{m^2+\sqrt{m^4+2\lambda\mu^4}}}\frac{d}{d\zeta}
	{\rm sn}(\zeta,\kappa)=\sqrt{\frac{2\mu^4}{m^2+\sqrt{m^4+2\lambda\mu^4}}}{\rm cn}(\zeta,\kappa){\rm dn}(\zeta,\kappa).
\end{equation}
For consistency reasons, we choose the phase $\chi'$ in such a way to have $G_2(0)=0$. This is obtained with the solutions of the equation ${\rm cn}(\chi',1)=0$ that are given by
\begin{equation}
   \chi'_k=(4k+1)K(-1)
\end{equation}
being $k$ an integer.
This fixes the arbitrary phase as required for the boundary conditions to be applied to Dyson-Schwinger equations. We note anyway that some arbitrariness remains.
Then, In the rest frame the propagator takes the simple form for $t>t'$
\begin{eqnarray}
   G_2({\bm x}-{\bm x'},t-t')&=&-\delta^3({\bm x}-{\bm x'})\frac{1}{(8\lambda)^\frac{1}{4}\mu}\theta(t-t'){\rm cn}\left((\lambda/2)^\frac{1}{4}\mu(t-t')+\chi'_k,-1\right)\times \nonumber \\
	&&{\rm dn}\left((\lambda/2)^\frac{1}{4}\mu(t-t')+\chi'_k,-1\right)
\end{eqnarray}
to which we have to add the similar contribution for $t<t'$. From this it is very easy to obtain the two-point function \cite{Frasca:2013tma}
\begin{equation}
   G_2(p)=\frac{\pi^3}{4K^3(-1)}
	\sum_{n=0}^\infty\frac{e^{-(n+\frac{1}{2})\pi}}{1+e^{-(2n+1)\pi}}(2n+1)^2\frac{1}{p^2-m_n^2+i\epsilon}
\end{equation}
given the mass spectrum
\begin{equation}
   m_n=(2n+1)\frac{\pi}{2K(-1)}\left(\frac{\lambda}{2}\right)^\frac{1}{4}\mu.
\end{equation}
This will solve the equation for $G_2$ provided we, consistently, will have in the following $G_3(0,x-y)=0$ and $G_4(0,0,x-y)=0$. Indeed, we have, after currents are set to zero,
\begin{eqnarray}
    &&\partial^2G_3(x-y,x-z)+\lambda\left[6G_1(x)G_2(x-y)G_2(x-z)+3G_1^2(x)G_3(x-y,x-z)\right. \\ \nonumber
    &&+3G_2(x-z)G_3(0,x-y)+3G_2(x-y)G_3(0,x-z) \\ \nonumber
    &&\left.+3G_2(0)G_3(x-y,x-z)+3G_1(x)G_4(0,x-y,x-z)+G_5(0,0,x-y,x-z)\right]=0 \\ \nonumber
    &&\\ \nonumber
    &&\partial^2G_4(x-y,x-z,x-w)+\lambda\left[6G_2(x-y)G_2(x-z)G_2(x-w)
    \right. \\ \nonumber
    &&+6G_1(x)G_2(x-y)G_3(x-z,x-w)+6G_1(x)G_2(x-z)G_3(x-y,x-w)\\ \nonumber
    &&+6G_1(x)G_2(x-w)G_3(x-y,x-z)+3G_1^2(x)G_4(x-y,x-z,x-w) \\ \nonumber
    &&+3G_2(x-y)G_4(0,x-z,x-w)+3G_2(x-z)G_4(0,x-y,x-w)  \\ \nonumber
    &&+3G_2(x-w)G_4(0,x-y,x-z)+3G_2(0)G_4(x-y,x-z,x-w) \\ \nonumber
    &&\left.+3G_1(x)G_5(0,x-y,x-z,x-w)+G_6(0,0,x-y,x-z,x-w)\right]=0 \\ \nonumber
    &\vdots&
\end{eqnarray}
that are solved by
\begin{equation}
\label{eq:G_3}
   G_3(x-y,x-z)=-6\lambda\int dx_1 G_2(x-x_1)G_1(x_1-y)G_2(x_1-y)G_2(x_1-z) 
\end{equation}
and it is easy to verify that $G_3(0,x-z)=G_3(x-y,0)=0$ using the property of Heaviside function $\theta(x)\theta(-x)=0$, and
\begin{eqnarray}
    &&G_4(x-y,x-z,x-w)=-6\lambda\int dx_1 G_2(x-x_1)G_2(x_1-y)G_2(x_1-z)G_2(x_1-w) \\ \nonumber
    &&-6\lambda\int dx_1G_2(x-x_1)\left[G_1(x_1-y)G_2(x_1-y)G_3(x_1-z,x_1-w)\right. \\ \nonumber
    &&\left.+G_1(x_1)G_2(x_1-z)G_3(x_1-y,x_1-w)
    +G_1(x_1-y)G_2(x_1-w)G_3(x_1-y,x_1-z)\right].
\end{eqnarray}
and it is to verify that $G_4(0,0,x-y)=0$. These hold provided that
\begin{eqnarray}
   G_4(0,x-y,x-z)&=&0 \\ \nonumber 
   G_5(0,0,x-y,x-z) &=& 0 
\end{eqnarray}
and
\begin{eqnarray}
    G_5(0,x-y,x-z,x-w)&=& 0 \\ \nonumber
    G_6(0,0,x-y,x-z,x-w)&=& 0.
\end{eqnarray}


\section{Renormalization conditions\label{sec2a}}

In order to get the exact solutions to the correlation functions, we have imposed a set of conditions on quantities that need renormalization. We work out here a couple of them to see how renormalization grants that what we have obtained is consistent. To get $G_1(x)$ we need to have $G_2(0)$ finite. This means that the integral
\begin{equation}
   I_2(d)=\frac{\pi^3}{4K^3(-1)}\int\frac{d^4p}{(2\pi)^4}
	\sum_{n=0}^\infty\frac{e^{-(n+\frac{1}{2})\pi}}{1+e^{-(2n+1)\pi}}(2n+1)^2\frac{1}{p^2-m_n^2+i\epsilon}
\end{equation}
must be regularized. 
To do this, we use dimensional regularization to obtain
\begin{equation}
   I_2(d)=\frac{\pi^3}{4K^3(-1)}\sum_{n=0}^\infty\frac{e^{-(n+\frac{1}{2})\pi}}{1+e^{-(2n+1)\pi}}(2n+1)^2\mu^{4-d}\int\frac{d^dp}{(2\pi)^d}
	\frac{1}{p^2-m_n^2+i\epsilon}
\end{equation}
where we have freely interchanged sum and integration. This yields
\begin{equation}
  I_2(\epsilon)=\frac{\pi^3}{4K^3(-1)}\sum_{n=0}^\infty\frac{e^{-(n+\frac{1}{2})\pi}}{1+e^{-(2n+1)\pi}}(2n+1)^2\frac{im_n^2}{(4\pi)^2}\left(\frac{1}{\epsilon}
	-\gamma-\ln\left(\frac{m_n^2}{4\pi\mu^2}\right)\right)
\end{equation}
being $\epsilon=4-d$, to be taken in the limit $\epsilon\rightarrow 0$. Now, turning back to the dispersion relation (\ref{eq:disp}), we realize that this divergent term can be absorbed into a renormalization of the coupling $\lambda$. This is so because
\begin{equation}
  I_2(\epsilon)=i\kappa_0\frac{\pi^3}{256K^5(-1)}\lambda^\frac{1}{2}\mu^2\frac{1}{\epsilon}
\end{equation}
with $\kappa_0=1.215018785\ldots$ arising from the evaluation of the sum. Indeed,
\begin{equation}
  m^2=3\lambda G_2(0)=\kappa_0\frac{3\pi^3}{256K^5(-1)}\lambda^\frac{3}{2}\mu^2\frac{1}{\epsilon}.
\end{equation}
In the small coupling limit, the propagator at the leading order is just $1/p^2$ when one notes that $\sum_nB_n=1$. The theory is free and no renormalization occurs at the leading order. Then, ordinary perturbation theory just follows.

The next step is to evaluate $G_3(0,0)$. Using eq.(\ref{eq:G_3}) one has
\begin{equation}
   G_3(0,0)=-6\lambda\int dx_1 G_2(x-x_1)G_1(x_1-x)G_2(x_1-x)G_2(x_1-x). 
\end{equation}
This integral is zero due to $G_2(x-x_1)G_2(x_1-x)=0$ that have opposite support. Higher order correlation functions can be treated similarly.

\section{Stability of the classical solutions\label{sec2b}}

One could ask if the exact solutions we started from to build the expansion for the scalar field are a real minimum for the functional of the field. This, in view of the mapping theorem proven in \cite{Frasca:2007uz,Frasca:2009yp}, will immediately apply to the Yang-Mills equations we exploit in the following sections. 
We will follow a different strategy that will take us to the same conclusions given in e.g. \cite{Goldstone:1974gf,Weinberg:2012pjx} that an eigenvalue problem must be solved to grant stability.
%
Firstly, let us see how the idea works for the free field. One has
\begin{equation}
    {\cal L}[\phi] = \int d^4x\left[\frac{1}{2}(\partial\phi)^2-\frac{1}{2}m^2\phi^2\right].
\end{equation}
For a given classical solution $\phi_0$, we can take a functional Taylor series of this as
\begin{equation}
    {\cal L} = {\cal L}[\phi_0]+\int d^4x'\left.\frac{\delta{\cal L}}{\delta\phi(x')}\right|_{\phi=\phi_0}\phi(x')
		+\frac{1}{2}\int d^4x'd^4x''\left.\frac{\delta^2{\cal L}}{\delta\phi(x')\delta\phi(x'')}\right|_{\phi=\phi_0}\phi(x')\phi(x'')+\ldots
\end{equation}
that becomes
\begin{equation}
    {\cal L} = {\cal L}[\phi_0]
		-\frac{1}{2}\int d^4x'd^4x\left[\partial^2\delta^4(x-x')+m^2\delta^4(x-x')\right]\phi(x')\phi(x)+\ldots,    
\end{equation}
where use has been made of the condition $\left.\frac{\delta{\cal L}}{\delta\phi(x')}\right|_{\phi=\phi_0}=0$, due to motion equation. We can now prove that the term originating from the second functional derivative is indeed strictly positive off-shell. Let us consider the eigenvalue problem
\begin{equation}
   \partial^2\chi_n+m^2\chi_n=\lambda_n\chi_n.
\end{equation}
Then, one has $\delta^4(x-x')=\sum_n\chi_n(x)\chi_n(x')$. It is not difficult to see that $\lambda_n=\lambda(p)=-(p^2-m^2)\le 0$ as the theory has a lower bound to energy. This means that
\begin{equation}
    {\cal L} = {\cal L}[\phi_0]+\frac{1}{2}\int d^4x'd^4x\sum_p(p^2-m^2)\chi_p(x)\chi_p(x')\phi(x')\phi(x)+\ldots.    
\end{equation}
But,
\begin{equation}
   c(p)=\int d^4x\chi_p(x)\phi(x)
\end{equation}
are the coefficients of the Fourier series for the field in terms of the eigenfunctions $\chi_p$ and then
\begin{equation}
    {\cal L} = {\cal L}[\phi_0]+\frac{1}{2}\sum_p(p^2-m^2)c^2(p)+\ldots={\cal L}[\phi_0]+\frac{1}{2}\sum_p(p^2-m^2)c^2(p)+\ldots.    
\end{equation}
The second term on the lhs must be positive definite for physical reasons and the classical solution is a minimum of the functional. For this it is enough to assume a lower bound on the spectrum for $p^2=m^2$ that grants the positivity of the energy. The zero mode is expected due to the translational invariance of the theory.

Then, we consider the action functional
\begin{equation}
    {\cal L}[\phi] = \int d^4x\left[\frac{1}{2}(\partial\phi)^2-\frac{\lambda}{4}\phi^4\right].
\end{equation}
Then,
\begin{equation}
    {\cal L} = {\cal L}[\phi_0]
		-\frac{1}{2}\int d^4x'd^4x\left[\partial^2\delta^4(x-x')+3\lambda\phi_0^2(x)\delta^4(x-x')\right]\phi(x')\phi(x)+\ldots    
\end{equation}
being now $\phi_0$ given by a solution to $\partial^2\phi_0+\lambda\phi_0^3=0$ (see \cite{Frasca:2009bc}). We introduce the set of eigenfunctions
\begin{equation}
    \partial^2\varphi_n+3\lambda\phi_0^2(x)\varphi_n=\epsilon_n\varphi_n.
\end{equation}
So, let us consider $\phi_0=\mu(2/\lambda)^\frac{1}{4}{\rm sn}(p\cdot x,-1)$ being sn a Jacobi elliptical function, then the eigenfunctions take the form
\begin{equation}
     \varphi_\mu=C\cdot{\rm sn}(p\cdot x,-1){\rm cn}(p\cdot x,-1)
\end{equation}
with eigenvalues $\epsilon(\mu)=-3\mu^2\sqrt{\lambda/2}\le 0$ with $\mu$ running from 0 to infinity. This holds on-shell and shows that 
our classical solutions are a minimum for the action functional, provided we work on-shell. We see again the zero mode showing that translational invariance applies for the theory.

This applies directly to Yang-Mills theory as we have shown a mapping theorem between the solutions of the scalar field theory and those of Yang-Mills theory\cite{Frasca:2009bc,Frasca:2007uz,Frasca:2009yp}. This theorem grants that, just in the Landau gauge, the mapping is exact.\footnote{The correctness of this theorem was agreed with T.~Tao after that a problem in the proof was properly fixed (see \url{http://wiki.math.toronto.edu/DispersiveWiki/index.php/Talk:Yang-Mills_equations).}}


\section{Classical Yang-Mills fields}
\label{sec3}

Our aim is to solve the classical equations of motion
\begin{equation}
\partial^\mu\partial_\mu A^a_\nu-\left(1-\frac{1}{\alpha}\right)\partial_\nu(\partial^\mu A^a_\mu)+gf^{abc}A^{b\mu}(\partial_\mu A^c_\nu-\partial_\nu A^c_\mu)+gf^{abc}\partial^\mu(A^b_\mu A^c_\nu)+g^2f^{abc}f^{cde}A^{b\mu}A^d_\mu A^e_\nu = j_\nu^a.
\end{equation}
To obtain the general solution to these equations we have to consider the equations with $j_\nu=0$. If we are able to solve them, we will be able to treat the full quantum theory. We specialize to SU(2) for the sake of simplicity, with $f^{abc}=\varepsilon_{abc}$ the Levi-Civita symbol, as
\begin{equation}
\partial^\mu\partial_\mu A^a_\nu-\left(1-\frac{1}{\alpha}\right)\partial_\nu(\partial^\mu A^a_\mu)+g\varepsilon_{abc}A^{b\mu}(\partial_\mu A^c_\nu-\partial_\nu A^c_\mu)+g\varepsilon_{abc}\partial^\mu(A^b_\mu A^c_\nu)+g^2\varepsilon_{abc}\varepsilon_{cde}A^{b\mu}A^d_\mu A^e_\nu = 0.
\end{equation}
Using the fundamental identity
\begin{equation}
\varepsilon_{ijk}\varepsilon_{imn} = \delta_{jm}\delta_{kn} - \delta_{jn}\delta_{km},
\end{equation}
one has
\begin{eqnarray}
&&\partial^\mu\partial_\mu A^a_\nu-\left(1-\frac{1}{\alpha}\right)\partial_\nu(\partial^\mu A^a_\mu)+g\varepsilon_{abc}A^{b\mu}(\partial_\mu A^c_\nu-\partial_\nu A^c_\mu)+g\varepsilon_{abc}\partial^\mu(A^b_\mu A^c_\nu) \nonumber \\
&&+g^2(A^{e\mu}A^a_\mu A^e_\nu-A^{b\mu}A^b_\mu A^a_\nu) = 0.
\end{eqnarray}
Now we take $A_\mu^a=(A_\mu^1,A_\mu^2,A_\mu^3)$ and zero for all other components. This gives
\begin{eqnarray}
&&\partial^\mu\partial_\mu A^1_\nu-\left(1-\frac{1}{\alpha}\right)\partial_\nu(\partial^\mu A^1_\mu)+ \nonumber \\
&&gA^{2\mu}(\partial_\mu A^3_\nu-\partial_\nu A^3_\mu)-gA^{3\mu}(\partial_\mu A^2_\nu-\partial_\nu A^2_\mu)+
g\partial^\mu(A^2_\mu A^3_\nu)-g\partial^\mu(A^3_\mu A^2_\nu)+ \nonumber \\
&&g^2(A^{2\mu}A^1_\mu A^2_\nu+A^{3\mu}A^1_\mu A^3_\nu-A^{2\mu}A^2_\mu A^1_\nu-A^{3\mu}A^3_\mu A^1_\nu) = 0 \nonumber \\
&&\partial^\mu\partial_\mu A^2_\nu-\left(1-\frac{1}{\alpha}\right)\partial_\nu(\partial^\mu A^2_\mu)+ \nonumber \\
&&gA^{3\mu}(\partial_\mu A^1_\nu-\partial_\nu A^1_\mu)-gA^{1\mu}(\partial_\mu A^3_\nu-\partial_\nu A^3_\mu)+
g\partial^\mu(A^3_\mu A^1_\nu)-g\partial^\mu(A^1_\mu A^3_\nu)+ \nonumber \\
&&g^2(A^{1\mu}A^2_\mu A^1_\nu+A^{3\mu}A^2_\mu A^3_\nu-A^{1\mu}A^1_\mu A^2_\nu-A^{3\mu}A^3_\mu A^2_\nu) = 0 \nonumber \\
&&\partial^\mu\partial_\mu A^3_\nu-\left(1-\frac{1}{\alpha}\right)\partial_\nu(\partial^\mu A^3_\mu)+ \nonumber \\
&&gA^{1\mu}(\partial_\mu A^2_\nu-\partial_\nu A^2_\mu)-gA^{2\mu}(\partial_\mu A^1_\nu-\partial_\nu A^1_\mu)+
g\partial^\mu(A^1_\mu A^2_\nu)-g\partial^\mu(A^2_\mu A^1_\nu)+ \nonumber \\
&&g^2(A^{1\mu}A^3_\mu A^1_\nu+A^{2\mu}A^3_\mu A^2_\nu-A^{1\mu}A^1_\mu A^3_\nu-A^{2\mu}A^2_\mu A^3_\nu) = 0.
\end{eqnarray}
Let us now put \cite{Smilga:2001ck} $A_1^1=A_2^2=A_3^3=\phi$ and all other components to zero. The set collapses on the equations
\begin{eqnarray}
&&\partial^2\phi-\left(1-\frac{1}{\alpha}\right)\partial_1(\partial^1\phi)+2g^2\phi^3 = 0 \nonumber \\
&&\partial^2\phi-\left(1-\frac{1}{\alpha}\right)\partial_2(\partial^2\phi)+2g^2\phi^3 = 0 \nonumber \\
&&\partial^2\phi-\left(1-\frac{1}{\alpha}\right)\partial_3(\partial^3\phi)+2g^2\phi^3 = 0
\end{eqnarray}
but this is possible only for $\alpha=1$. This shows how the choice of the Landau gauge simplifies computations. But we can also take $A_1^1\ne A_2^2\ne A_3^3\ne 0$. This will yield the following set of equations
\begin{eqnarray}
&&\partial^2A^1_1-\left(1-\frac{1}{\alpha}\right)\partial_1(\partial^1 A^1_1)+g^2[(A^2_2)^2+(A^3_3)^2] A^1_1 = 0 \nonumber \\
&&\partial^2A^2_2-\left(1-\frac{1}{\alpha}\right)\partial_2(\partial^2 A^2_2)+g^2[(A^1_1)^2+(A^3_3)^2] A^2_2 = 0 \nonumber \\
&&\partial^2A^3_3-\left(1-\frac{1}{\alpha}\right)\partial_3(\partial^3 A^3_3)+g^2[(A^1_1)^2+(A^2_2)^2] A^3_3 = 0.
\end{eqnarray}
We have shown that in this case the solutions change from exact to asymptotic for $g\rightarrow\infty$ \cite{Frasca:2009yp}. We would like to follow a different approach and get the exact solution. So, given the {\sl ansatz}
\begin{eqnarray}
A^{1}_1&=&X\cdot{\rm sn}(p\cdot x+\chi,-1) \nonumber \\
A^{2}_2&=&Y\cdot{\rm sn}(p\cdot x+\chi,-1) \nonumber \\ 
A^{3}_3&=&Z\cdot{\rm sn}(p\cdot x+\chi,-1)
\end{eqnarray}
with $\chi$ an arbitrary phase, the dispersion relation $p^2=\mu^2g$ to hold and $\mu$ an integration constant with the dimension of an energy, we get the following set of algebraic equations
\begin{eqnarray}
    Y^2+Z^2&=&\frac{2}{g^2}\left(1-\frac{1}{\alpha}\right)p_1^2+\mu^2\frac{2}{g} \nonumber \\
		X^2+Z^2&=&\frac{2}{g^2}\left(1-\frac{1}{\alpha}\right)p_2^2+\mu^2\frac{2}{g} \nonumber \\
		X^2+Y^2&=&\frac{2}{g^2}\left(1-\frac{1}{\alpha}\right)p_3^2+\mu^2\frac{2}{g}
\end{eqnarray}
that is easily solved. This shows that the idea of an asymptotic mapping in \cite{Frasca:2009yp} was correct as the contributions that select the different components goes like $O(1/g^2)$ and so negligible in the limit $g\rightarrow\infty$. But here we proved that such a mapping is indeed exact provided the proper solutions for the given gauge are used. For $\alpha=1$ (Landau gauge) one has the expected result \cite{Frasca:2009yp}
\begin{equation}
\label{eq:exsol}
   A_1^1=A_2^2=A_3^3=\frac{\mu}{g^\frac{1}{2}}\cdot{\rm sn}(p\cdot x,-1).
\end{equation}
These are the solutions we will adopt in the following as we fix the gauge to be Landau granting simpler calculations.

\section{Dyson-Schwinger equations for Yang-Mills theory}
\label{sec4}

In order to derive the Dyson-Schwinger equations for a Yang-Mills theory with the technique seen for the scalar field we need to account also for the ghost field of the theory. The quantum equations of motion take the form
\begin{eqnarray}
   &&\partial^\mu\partial_\mu A^a_\nu+gf^{abc}A^{b\mu}(\partial_\mu A^c_\nu-\partial_\nu A^c_\mu)+gf^{abc}\partial^\mu(A^b_\mu A^c_\nu)+g^2f^{abc}f^{cde}A^{b\mu}A^d_\mu A^e_\nu \nonumber \\
	&&= gf^{abc}\partial_\nu(\bar c^b c^c) + j_\nu^a \nonumber \\
	 &&\partial^\mu\partial_\mu c^a+gf^{abc}\partial^\mu(A_\mu^bc^c)=\varepsilon^a 
\end{eqnarray}
Here we have taken the Landau gauge, $\alpha=1$, and $c,\ \bar c$ are the ghost fields. Averaging on the vacuum state and dividing by the partition function $Z_{YM}[j,\bar\varepsilon,\varepsilon]$ one has
\begin{eqnarray}
    &&\partial^2G_{1\nu}^{(j)a}(x)+gf^{abc}(\langle A^{b\mu}\partial_\mu A^c_\nu\rangle-\langle A^{b\mu}\partial_\nu A^c_\mu\rangle)Z^{-1}_{YM}[j,\bar\varepsilon,\epsilon]
		+gf^{abc}\partial^\mu\langle A^b_\mu A^c_\nu\rangle Z^{-1}_{YM}[j,\bar\varepsilon,\varepsilon]
		\nonumber \\
		&&+g^2f^{abc}f^{cde}\langle A^{b\mu}A^d_\mu A^e_\nu\rangle Z^{-1}_{YM}[j,\bar\varepsilon,\varepsilon] 
		=gf^{abc}\langle\partial_\nu(\bar c^b c^c)\rangle Z^{-1}_{YM}[j,\bar\varepsilon,\varepsilon] + j_\nu^a \nonumber \\
	 &&\partial^2 P^{(\varepsilon)a}_1(x)+gf^{abc}\partial^\mu\langle A_\mu^bc^c\rangle Z^{-1}_{YM}[j,\bar\varepsilon,\varepsilon]=\varepsilon^a 
\end{eqnarray}
We have introduced the one-point functions
\begin{eqnarray}
    &&G_{1\nu}^{(j)a}(x)Z_{YM}[j,\bar\varepsilon,\epsilon]=\langle A^a_\nu(x)\rangle \nonumber \\
		&&P^{(\varepsilon)a}_1(x)Z_{YM}[j,\bar\varepsilon,\epsilon]=\langle c^a(x)\rangle 
\end{eqnarray}
We derive once with respect to currents obtaining
\begin{eqnarray}
   &&G_{2\nu\kappa}^{(j)ab}(x,x)Z_{YM}[j,\bar\varepsilon,\epsilon]+G_{1\nu}^{(j)a}(x)G_{1\kappa}^{(j)b}(x)Z_{YM}[j,\bar\varepsilon,\epsilon]=\langle A^a_\nu(x)A^b_\kappa(x)\rangle \nonumber \\
	 &&P^{(\varepsilon)ab}_2(x,x)Z_{YM}[j,\bar\varepsilon,\epsilon]+\bar P^{(\varepsilon)a}_1(x)P^{(\varepsilon)b}_1(x)Z_{YM}[j,\bar\varepsilon,\epsilon]=
	\langle \bar c^b(x)c^a(x)\rangle \nonumber \\
	&&\partial_\mu G_{2\nu\kappa}^{(j)ab}(x,x)Z_{YM}[j,\bar\varepsilon,\epsilon]+\partial_\mu G_{1\nu}^{(j)a}(x)G_{1\kappa}^{(j)b}(x)Z_{YM}[j,\bar\varepsilon,\epsilon]=
	\langle\partial_\mu A^a_\nu(x)A^b_\kappa(x)\rangle \nonumber \\
	&&K^{(\varepsilon,j)ab}_{2\nu}(x,x)Z_{YM}[j,\bar\varepsilon,\epsilon]+P^{(\varepsilon)a}_1(x)G_{1\nu}^{(j)b}(x)Z_{YM}[j,\bar\varepsilon,\epsilon]=\langle c^a(x)A_\nu^b(x)\rangle
\end{eqnarray}
and twice to obtain
\begin{eqnarray}
   &&G_{3\nu\kappa\rho}^{(j)abc}(x,x,x)Z_{YM}[j,\bar\varepsilon,\epsilon]+G_{2\nu\kappa}^{(j)ab}(x,x)G_{1\rho}^{(j)c}(x)Z_{YM}[j,\bar\varepsilon,\epsilon]+ \nonumber \\
	&&G_{2\nu\rho}^{(j)ac}(x,x)G_{1\kappa}^{(j)b}(x)Z_{YM}[j,\bar\varepsilon,\epsilon]
	+G_{1\nu}^{(j)a}(x)G_{2\kappa\rho}^{(j)bc}(x,x)Z_{YM}[j,\bar\varepsilon,\epsilon]+ \nonumber \\
	&&G_{1\nu}^{(j)a}(x)G_{1\kappa}^{(j)b}(x)G_{1\rho}^{(j)c}(x)Z_{YM}[j,\bar\varepsilon,\epsilon]=\langle A^a_\nu(x)A^b_\kappa(x)A^c_\rho(x)\rangle.
\end{eqnarray}
So, we get the first set of Dyson-Schwinger equations as it is
\begin{eqnarray}
\label{eq:ds_1}
    &&\partial^2G_{1\nu}^{(j)a}(x)+gf^{abc}(
		\partial^\mu G_{2\mu\nu}^{(j)bc}(x,x)+\partial^\mu G_{1\mu}^{(j)b}(x)G_{1\nu}^{(j)c}(x)-
		\partial_\nu G_{2\mu}^{(j)\mu bc}(x,x)-\partial_\nu G_{1\mu}^{(j)b}(x)G_{1}^{(j)\mu c}(x)) \nonumber \\
		&&+gf^{abc}\partial^\mu G_{2\mu\nu}^{(j)bc}(x,x)+gf^{abc}\partial^\mu(G_{1\mu}^{(j)b}(x)G_{1\nu}^{(j)c}(x))		
		\nonumber \\
		&&+g^2f^{abc}f^{cde}(G_{3\mu\nu}^{(j)\mu bde}(x,x,x)+G_{2\mu\nu}^{(j)bd}(x,x)G_{1}^{(j)\mu e}(x) \nonumber \\
	&&+G_{2\nu\rho}^{(j)ac}(x,x)G_{1}^{(j)\rho b}(x)
	+G_{1}^{(j)\mu b}(x)G_{2\mu\nu}^{(j)de}(x,x)+ \nonumber \\
	&&G_{1}^{(j)\mu b}(x)G_{1\mu}^{(j)d}(x)G_{1\nu}^{(j)e}(x))
		=gf^{abc}(\partial_\nu P^{(\varepsilon)bc}_2(x,x)+\partial_\nu (\bar P^{(\varepsilon)b}_1(x)P^{(\varepsilon)c}_1(x))) + j_\nu^a \nonumber \\
	 &&\partial^2 P^{(\varepsilon)a}_1(x)+gf^{abc}\partial^\mu
	(K^{(\varepsilon,j)bc}_{2\mu}(x,x)+P^{(\varepsilon)b}_1(x)G_{1\mu}^{(j)c}(x))=\varepsilon^a. 
\end{eqnarray}
Now we put the currents to zero and noticing that by translation invariance is $G_2(x,x)=G_2(x-x)=G_2(0)$, $G_3(x,x,x)=G_3(0,0)$ and $K_2(x,x)=K_2(0)$ one has
\begin{eqnarray}
    &&\partial^2G_{1\nu}^{a}(x)+gf^{abc}(
		\partial^\mu G_{2\mu\nu}^{bc}(0)+\partial^\mu G_{1\mu}^{b}(x)G_{1\nu}^{c}(x)-
		\partial_\nu G_{2\mu}^{\nu bc}(0)-\partial_\nu G_{1\mu}^{b}(x)G_{1}^{\mu c}(x)) \nonumber \\
		&&+gf^{abc}\partial^\mu G_{2\mu\nu}^{bc}(0)+gf^{abc}\partial^\mu(G_{1\mu}^{b}(x)G_{1\nu}^{c}(x))		
		\nonumber \\
		&&+g^2f^{abc}f^{cde}(G_{3\mu\nu}^{\mu bde}(0,0)+G_{2\mu\nu}^{bd}(0)G_{1}^{\mu e}(x) \nonumber \\
	&&+G_{2\nu\rho}^{ac}(0)G_{1}^{\rho b}(x)
	+G_{1}^{\mu b}(x)G_{2\mu\nu}^{de}(0)+ \nonumber \\
	&&G_{1}^{\mu b}(x)G_{1\mu}^{d}(x)G_{1\nu}^{e}(x))
		=gf^{abc}(\partial_\nu P^{bc}_2(0)+\partial_\nu (\bar P^{b}_1(x)P^{c}_1(x))) \nonumber \\
	 &&\partial^2 P^{a}_1(x)+gf^{abc}\partial^\mu
	(K^{bc}_{2\mu}(0)+P^{b}_1(x)G_{1\mu}^{c}(x))=0. 
\end{eqnarray}
This set of Dyson-Schwinger equations is exactly solved by the exact solutions in the Landau gauge provided in Sec.~\ref{sec3} with the choice of the conditions $G_{2\mu\nu}^{ab}(0)=0$, $P_2^{ab}(0)=0$, $G_{3\mu\nu}^{\mu bde}(0,0)=0$ and $K^{bc}_{2\mu}(0)=0$. In such a case, it is immediately seen that the ghost one-point function decouples and can be taken constant. So, it does not contribute to the gluon one-point function.

The Dyson-Schwinger equation for the two-point functions can be obtained by further deriving eq.(\ref{eq:ds_1}). We get
\begin{eqnarray}
\label{eq:ds_2}
    &&\partial^2G_{2\nu\kappa}^{(j)am}(x-y)+gf^{abc}(
		\partial^\mu G_{3\mu\nu\kappa}^{(j)bcm}(x,x,y)+\partial^\mu G_{2\mu\kappa}^{(j)bm}(x-y)G_{1\nu}^{(j)c}(x)
		+\partial^\mu G_{1\mu}^{(j)b}(x)G_{2\nu\kappa}^{(j)cm}(x-y) \nonumber \\
		&&-\partial_\nu G_{3\mu\kappa}^{(j)\mu bcm}(x,x,y)-\partial_\nu G_{2\mu\kappa}^{(j)bm}(x-y)G_{1}^{(j)\mu c}(x)) 
		-\partial_\nu G_{1\mu}^{(j)b}(x)G_{2\kappa}^{(j)\mu cm}(x-y))
		\nonumber \\
		&&+gf^{abc}\partial^\mu G_{3\mu\nu\kappa}^{(j)bcm}(x,x,y)+gf^{abc}\partial^\mu(G_{2\mu\kappa}^{(j)bm}(x-y)G_{1\nu}^{(j)c}(x))
				+gf^{abc}\partial^\mu(G_{1\mu}^{(j)b}(x)G_{1\nu\kappa}^{(j)cm}(x-y))
		\nonumber \\
		&&+g^2f^{abc}f^{cde}(G_{4\mu\nu\kappa}^{(j)\mu bdem}(x,x,x,y)+G_{3\mu\nu\kappa}^{(j)bdm}(x,x,y)G_{1}^{(j)\mu e}(x) 
		+G_{2\mu\nu}^{(j)bd}(x,x)G_{2\kappa}^{(j)\mu em}(x-y)\nonumber \\
	&&+G_{3\nu\rho\kappa}^{(j)acm}(x,x,y)G_{1}^{(j)\rho b}(x)+G_{2\nu\rho}^{(j)ac}(x,x)G_{2\kappa}^{(j)\rho b}(x-y)
	+G_{1}^{(j)\mu b}(x)G_{3\mu\nu\kappa}^{(j)dem}(x,x,y)+ \nonumber \\
	&&G_{2\kappa}^{(j)\mu bm}(x-y)G_{1\mu}^{(j)d}(x)G_{1\nu}^{(j)e}(x)+
	G_{1}^{(j)\mu b}(x)G_{2\mu\kappa}^{(j)dm}(x-y)G_{1\nu}^{(j)e}(x)+
	G_{1}^{(j)\mu b}(x)G_{1\mu}^{(j)d}(x)G_{2\nu\kappa}^{(j)em}(x-y)) \nonumber \\
	&&	=gf^{abc}(\partial_\nu K^{(j\varepsilon)bcm}_{3\kappa}(x,x,y)+\partial_\nu (\bar P^{(\varepsilon)b}_1(x)K^{(j\varepsilon)cm}_{2\kappa}(x,y))) 
	+\partial_\nu (\bar K^{(j\varepsilon)bm}_{2\kappa}(x,y)P^{(\varepsilon)c}_1(x)))
	+ \delta_{ab}g_{\nu\kappa}\delta^4(x-y) \nonumber \\
	 &&\partial^2 P^{(\varepsilon)am}_2(x-y)+gf^{abc}\partial^\mu
	(K^{(\varepsilon,j)bcm}_{3\mu}(x,x,y)+P^{(\varepsilon)bm}_2(x-y)G_{1\mu}^{(j)c}(x)+ \nonumber \\
	&&P^{(\varepsilon)b}_1(x)K_{2\mu}^{(j\varepsilon)cm}(x-y))=\delta_{am}\delta^4(x-y) \nonumber \\
	&&\partial^2 K^{(j\varepsilon)am}_{2\kappa}(x-y)+gf^{abc}\partial^\mu
	(L^{(\varepsilon,j)bcm}_{2\mu\kappa}(x,x,y)+ \nonumber \\
	&&K^{(j\varepsilon)bm}_{2\kappa}(x-y)G_{1\mu}^{(j)c}(x)+P^{(\varepsilon)b}_1(x)G_{2\mu\kappa}^{(j)cm}(x-y))=0. 
\end{eqnarray}
These become, setting currents to zero and using translation invariance,
\begin{eqnarray}
\label{eq:ds_3}
    &&\partial^2G_{2\nu\kappa}^{am}(x-y)+gf^{abc}(
		\partial^\mu G_{3\mu\nu\kappa}^{bcm}(0,x-y)+\partial^\mu G_{2\mu\kappa}^{bm}(x-y)G_{1\nu}^{c}(x)
		+\partial^\mu G_{1\mu}^{b}(x)G_{2\nu\kappa}^{cm}(x-y) \nonumber \\
		&&-\partial_\nu G_{3\mu\kappa}^{\mu bcm}(0,x-y)-\partial_\nu G_{2\mu\kappa}^{bm}(x-y)G_{1}^{\mu c}(x)) 
		-\partial_\nu G_{1\mu}^{b}(x)G_{2\kappa}^{\mu cm}(x-y))
		\nonumber \\
		&&+gf^{abc}\partial^\mu G_{3\mu\nu\kappa}^{bcm}(0,x-y)+gf^{abc}\partial^\mu(G_{2\mu\kappa}^{bm}(x-y)G_{1\nu}^{c}(x))
				+gf^{abc}\partial^\mu(G_{1\mu}^{b}(x)G_{1\nu\kappa}^{cm}(x-y))
		\nonumber \\
		&&+g^2f^{abc}f^{cde}(G_{4\mu\nu\kappa}^{\mu bdem}(0,0,x-y)+G_{3\mu\nu\kappa}^{bdm}(0,x-y)G_{1}^{\mu e}(x) 
		+G_{2\mu\nu}^{bd}(0)G_{2\kappa}^{\mu em}(x-y)\nonumber \\
	&&+G_{3\nu\rho\kappa}^{acm}(0,x-y)G_{1}^{\rho b}(x)+G_{2\nu\rho}^{ac}(0)G_{2\kappa}^{\rho b}(x-y)
	+G_{1}^{\mu b}(x)G_{3\mu\nu\kappa}^{dem}(0,x-y)+ \nonumber \\
	&&G_{2\kappa}^{\mu bm}(x-y)G_{1\mu}^{d}(x)G_{1\nu}^{e}(x)+
	G_{1}^{\mu b}(x)G_{2\mu\kappa}^{dm}(x-y)G_{1\nu}^{e}(x)+
	G_{1}^{\mu b}(x)G_{1\mu}^{d}(x)G_{2\nu\kappa}^{em}(x-y)) \nonumber \\
	&&	=gf^{abc}(\partial_\nu K^{bcm}_{3\kappa}(0,x-y)+\partial_\nu (\bar P^{b}_1(x)K^{cm}_{2\kappa}(x-y))) 
	+\partial_\nu (\bar K^{bm}_{2\kappa}(x-y)P^{c}_1(x)))
	+ \delta_{ab}g_{\nu\kappa}\delta^4(x-y) \nonumber \\
	 &&\partial^2 P^{am}_2(x-y)+gf^{abc}\partial^\mu
	(K^{bcm}_{3\mu}(0,x-y)+P^{bm}_2(x-y)G_{1\mu}^{c}(x)+ \nonumber \\
	&&P^{b}_1(x)K_{2\mu}^{cm}(x-y))=\delta_{am}\delta^4(x-y) \nonumber \\
	&&\partial^2 K^{am}_{2\kappa}(x-y)+gf^{abc}\partial^\mu
	(L^{bcm}_{2\mu\kappa}(0,x-y)+ \nonumber \\
	&&K^{bm}_{2\kappa}(x-y)G_{1\mu}^{c}(x)+P^{b}_1(x)G_{2\mu\kappa}^{cm}(x-y))=0. 
\end{eqnarray} eq.(\ref{eq:exsol}) 
Now we note that $G_{1\mu}^a(x)$ can be written as
\begin{equation}
   G_{1\mu}^a(x)=\eta^a_\mu\phi(x),
\end{equation}
being $\phi(x)=\mu\left(\frac{2}{Ng^2}\right)^\frac{1}{4}\cdot{\rm sn}(p\cdot x,-1)$ with $\eta^a_\mu$ constants and $p^2=\mu^2\sqrt{Ng^2/2}$. Then, 
this set of equations can be solved exactly. For the two-point function in the Landau gauge we can write
\begin{equation}
    G^{ab}_{\mu\nu}(x)=\delta_{ab}\left(g_{\mu\nu}-\frac{p_\mu p_\nu}{p^2}\right)\Delta(x-y)
\end{equation}
provided
\begin{eqnarray}
    &&\partial^2\Delta(x-y)+3Ng^2\phi^2(x)\Delta(x-y)=\delta^4(x-y) \nonumber \\
		&&P_1^a(x)=0 \nonumber \\
	 &&\partial^2 P^{am}_2(x-y)=\delta_{am}\delta^4(x-y) \nonumber \\
		&&K^{am}_{2\kappa}(x-y)=0. 
\end{eqnarray}
and $G_{2\nu\rho}^{ac}(0)=0$, $G_{3\mu\nu\kappa}^{bcm}(0,x-y)=0$, $G_{4\mu\nu\kappa}^{\mu bdem}(0,0,x-y)$, $K^{bcm}_{3\kappa}(0,x-y)=0$. This is the main result of the paper showing that the set of Dyson-Schwinger equations for Yang-Mills theory can be exactly solved at least to the level of two-point function. At this stage the results agree with lattice computations. The solutions are identical to the case of the scalar field theory provided one takes $\lambda=Ng^2$.

\section{Conclusions}
\label{sec5}

We solved exactly the Dyson-Schwinger set of equations for the scalar field and the Yang-Mills theory stopping to the two-point function,
provided a proper set of exact solutions of the classical equations of motion is selected
. These solutions display the interesting feature of a mass gap notwithstanding both the theories are massless. Built on the single particle state of the theory there is a set of excited states. Anyhow, it appears really interesting the possibility to treat nonlinear quantum field theory in such exact manner even if the theories appear treatable just by perturbative methods. We hope to extend this approach to other theories mainly to study their infrared behavior.




\end{document}